\documentclass[12pt]{iopart}

\newcommand{\eqref}[1]{(\ref{#1})}

\usepackage{iopams}
\usepackage{graphicx}
\usepackage{amsfonts}
\usepackage{amsbsy}
\usepackage{amssymb}

\usepackage[top=2cm,bottom=2cm,left=1.5cm,right=1.5cm,columnsep=0.7cm]{geometry}

\usepackage[T1]{fontenc}
\usepackage[utf8]{inputenc}
\usepackage{enumerate}   
\usepackage{bm}

\usepackage{microtype}

\usepackage{color}

\usepackage{array}

\usepackage{citesort}
\newcommand{\bu}{{\bf u}}

\newcommand{\bn}{{\bf n}}
\newcommand{\mb}{{\bf m}}

\begin{document}

\title{Modeling chiral active particles: from circular motion to odd interactions}

\author{Lorenzo Caprini$^1$}

\address{$^1$ Sapienza University of Rome, P.le Aldo Moro 5, IT-00185 Rome, Italy}

\ead{ lorenzo.caprini@uniroma1.it}

\author{Alessandro Petrini$^1$}

\address{$^1$ Sapienza University of Rome, P.le Aldo Moro 5, IT-00185 Rome, Italy}

\author{Umberto Marini Bettolo Marconi$^2$}

\address{$2$ Scuola di Scienze e Tecnologie, Universit\`a di Camerino - via Madonna delle Carceri, 62032, Camerino, Italy}


\vspace{10pt}
\begin{indented}
\item[]September 2025
\end{indented}

\begin{abstract}
In this paper, we discuss microscopic models for chiral active particles, i.e., rotating active units that exhibit circular or spinning motion.
While non-chiral active particles are typically governed by self-propulsion and conservative interactions, the rotating motion of chiral particles generates additional non-conservative forces that cannot be derived from a potential. These manifest as effective transverse forces, acting perpendicular to the line connecting the centres of two interacting particles, and are referred to as odd interactions, because they break the mirror symmetry of the system. Here, we demonstrate that odd interactions arise from a limiting case of a well-established model describing spinning granular objects. 
In addition, we show that these models for chiral active objects give rise to a novel collective phenomenon that emerges uniquely from transverse forces and, hence, chirality.
Specifically, the system undergoes a transition from a homogeneous phase to an inhomogeneous one characterised by regions depleted of particles, referred to as bubbles. This collective behaviour, termed BIO (bubbles induced by odd interactions), is a general emergent phenomenon arising from chirality and odd interactions.
In this work, we review theoretical approaches to this problem, including a scaling argument and predictions for spatial velocity correlations that account for the BIO phase. Finally, we outline perspectives and open challenges concerning this collective phenomenon.

\end{abstract}


\maketitle

\section{Introduction}

Chirality is the property of an object being non-superimposable on its mirror image and is a defining feature of a broad class of active matter systems~\cite{marchetti2013hydrodynamics,elgeti2015physics,bechinger2016active}. Chirality may arise from the particle's shape, allowing one to distinguish between left- and right-handed active units, or from the mechanism of activity itself, i.e., the ability to self-propel through a rotationally asymmetric process that converts energy from the environment or  an internal energy reservoir. This chiral activity~\cite{lowen2016chirality,liebchen2022chiral} gives rise to circular, helical, or spinning trajectories.
For example, several biological micro-swimmers including bacteria and spermatozoa possess intrinsic chirality, which enables them to follow spiral-like swimming paths or circular trajectories near surfaces~\cite{woolley2003motility}, and even to self-organise into rotating crystalline structures~\cite{petroff2015fast}. Similar chiral motion also appears in more complex biological systems, such as starfish embryos~\cite{tan2022odd} and isotropic droplets~\cite{carenza2019rotation,sevc2012geometrical}.
Chirality has also been engineered in the motion of artificial micro-swimmers, such as L-shaped self-propelled colloids~\cite{kummel2013circular} and spinning colloids actuated by magnetic fields~\cite{yan2015rotating,massana2021arrested}.
At larger scales, rotating trajectories have likewise been observed in macroscopic active units. Examples include circular walkers powered by light~\cite{siebers2023exploiting}, hexbug particles driven by internal motors~\cite{carrillo2025depinning}, spinning granular units sustained by airflow~\cite{vega2022diffusive,lopez2022chirality,farhadi2018dynamics,lopez2023spin}, and chiral active vibrobots on vibrating plates~\cite{scholz2018rotating,scholz2021surfactants,caprini2024spontaneous}.

The above examples illustrate that chiral active systems are ubiquitous, motivating extensive theoretical efforts to reproduce self-propelled circular motion.
At the single-particle level, a chiral active particle is characterised by a reduced long-time diffusion coefficient~\cite{van2008dynamics,caprini2019active,sevilla2016diffusion,li2020diffusion,khatri2022diffusion,van2022role,olsen2024optimal} and by an oscillating mean-square displacement~\cite{van2008dynamics,caprini2019active}, both consequences of circular trajectories. Moreover, chirality has been identified as a crucial ingredient underlying odd diffusivity~\cite{hargus2021odd,vega2022diffusive,kalz2022collisions,yasuda2022time,kalz2024oscillatory,hargus2024flux}, i.e., the emergence of fluxes oriented orthogonally to the direction of motion.
Single-particle theories have further revealed that chirality suppresses the wall accumulation typical of non-chiral active systems~\cite{caprini2019active,van2025three}, instead producing directed currents along walls~\cite{caprini2019active} or circulating edge currents in the presence of confining potentials~\cite{caprini2023chiral}.
At the collective scale, chirality drives a variety of novel phenomena, most of which have been investigated in systems governed by conservative interparticle interactions. On one hand, chirality is known to suppress motility-induced phase separation (MIPS), which is otherwise observed in non-chiral active systems~\cite{liao2018clustering,bickmann2022analytical}. This suppression gives rise to micro-phase separation~\cite{semwal2024macro}, clustering inhibition~\cite{ma2022dynamical,kreienkamp2022clustering}, and even hyperuniform states~\cite{lei2019nonequilibrium,zhang2022hyperuniform,kuroda2023microscopic}. On the other hand, the interplay between chirality and conservative forces generates a wide range of emergent behaviors~\cite{reichhardt2019reversibility,zhang2020reconfigurable,caprini2024self}, including self-reversing vorticity in cohesive clusters~\cite{caprini2024self}, spontaneous vortices~\cite{shee2024emergent,marconi2025spontaneous}, and angular momentum in ideal active crystals~\cite{furthauer2012active,marconi2025spontaneous}, with similar effects also reported for anisotropic interactions~\cite{lei2023collective}.
Finally, chirality has been shown to underlie peculiar caging effects in active glasses~\cite{debets2023glassy} and to strongly influence flocking behaviour in systems with alignment interactions~\cite{levis2018micro,levis2019simultaneous,huang2020dynamical,negi2023geometry}.

Chiral active systems are often described using coarse-grained macroscopic approaches such as hydrodynamic or elastodynamic theories characterised by odd coefficients~\cite{fruchart2023odd}, including odd viscosity~\cite{banerjee2017odd,markovich2021odd,lou2022odd,reichhardt2022active,hosaka2023lorentz,everts2024dissipative} and odd elasticity~\cite{scheibner2020odd,braverman2021topological,alexander2021layered,ishimoto2023odd,surowka2023odd}. 
At a coarse-grained level, chiral liquids
and chiral crystals are  described in terms of non symmetric viscosity  and non symmetric elastic tensors, respectively. Such theories not only reproduce experimental observations,  such as edge currents~\cite{van2016spatiotemporal}, but also predict unexpected phenomena, including transverse diffusion and oscillatory behaviour in the overdamped regime~\cite{scheibner2020odd}.
However, odd properties in chiral fluids and crystals do not arise from conservative interactions alone, as in the case of non-rotating particles, and require  different microscopic mechanisms.

In this paper, we examine microscopic models and experimental realizations of chiral active systems (Sec.~\ref{sec:models}) in which particles interact through effective transverse forces, i.e. acting in a direction perpendicular to the line joining their centres. 
Such forces cannot be represented as
gradients of a potential and they violate the principle of energy conservation. 
Furthermore, the Newtonian action-reaction principle is only weakly satisfied: although the particles exert equal and opposite forces on each other (i.e. the forces are reciprocal), the torques they exert are not equal and opposite.

These interactions are also named odd~\cite{caprini2025bubble}  because they break the mirror symmetry, i.e. do not satisfy the parity invariance. We show that such interactions are theoretically connected to granular spinners subject to rotational friction. While particle systems with odd interactions have only recently begun to be explored in simulations, initial studies in the over-damped regime with additional Lennard-Jones attractions revealed phase separation accompanied by edge currents at the cluster surface~\cite{caporusso2024phase}.
Here, by contrast, we focus on emergent collective phenomena that arise when the dynamics includes a small but finite amount of inertia. In particular, Ref.~\cite{caprini2025bubble} demonstrated that in the regime of large chirality, odd interactions drive a transition from a homogeneous to an inhomogeneous phase characterised by ordered bubbles. This phenomenon termed BIO (bubbles induced by odd interactions) occurs even in the absence of attractive forces (Sec.~\ref{sec:BIO}).
This work reviews numerical results and theoretical approaches that account for the BIO phase, ranging from scaling arguments clarifying the mechanism of bubble stability to predictions of spatial velocity correlations in high-density systems with solid-like order~\cite{caprini2025odd}. Finally, we outline open questions and future perspectives on chiral active systems governed by odd (transverse) interactions (Sec.~\ref{sec:conclusion}).


\section{Microscopic models for chiral active particles}\label{sec:models}

\subsection{Circular self-propelled motion}

A broad class of active particles exhibits chiral self-propelled motion, meaning that their trajectories take circular shapes in two dimensions and helical shapes in three dimensions.
The dynamics of an active chiral particle can be described by an equation that incorporates its persistence time 
$\tau$, typical self-propulsion speed $v_0$, and
 characteristic angular velocity $\omega$.
The particle’s trajectory is governed by the following underdamped equation of motion for its velocity, $\mathbf{v}=\dot{\mathbf{x}}$, governs the particle's trajectory:
\begin{equation}
\label{eq:translationaldynamics}
m\dot{\mathbf{v}} =-\gamma \mathbf{v} + \gamma\sqrt{2 D_t} \boldsymbol{\xi} + \gamma v_0 \mathbf{n} 
\end{equation}
where  $m$  is the mass and $\boldsymbol{\xi}$ is a white noise with unit variance and zero average. The constants $\gamma$ and $D_t$ represent the friction and translational diffusion coefficient, originating from the solvent (as in the case of colloids) or to any randomness in the propulsion mechanisms in the case of granular spinners. In the latter case, these constants do not satisfy the Einstein relation, or in other words, $D_t$ is completely independent of the environmental temperature.
The last force term $\gamma v_0 \mathbf{n}$ represents the self-propulsion force, causing the active motion with a typical speed $v_0$. This motion is directed along the orientational vector $\mathbf{n}$, whose time evolution depends on the specific model considered.
Hereafter, we specialize to two-dimensional systems.

\vskip10pt
\noindent
{\it Chiral active Brownian particles} -- According to this model \cite{van2008dynamics},  $\mathbf{n}$ is unit vector, with components $\mathbf{n}=(\cos(\theta), \sin(\theta))$, where $\theta$ is usually termed orientational angle. In this case, it is convenient to express the dynamics of $\mathbf{n}$ in polar coordinates, since $|\mathbf{n}|=1$, so that the dynamics read
\begin{equation}
\label{eq:chiralABP}
\dot{\theta}=\sqrt{2D_r} \eta + \Omega \,,
\end{equation}
where $\eta$ is a white noise with zero average and unit variance and $D_r$ corresponds to the particle rotational diffusion coefficient.
The constant $\Omega$ is a clockwise or counterclockwise drift in the angular dynamics. As a consequence, this term is responsible for circular trajectories for non-vanishing $\omega$, and thus can be identified as the particle chirality.
Even if in macroscopic descriptions (active granular particles), the dynamics for the orientational angle typically involves a rotational inertial term proportional to the moment of inertia; here, we neglect this contribution for simplicity.

\vskip10pt
\noindent
{\it Chiral active Ornstein-Uhlenbeck particles} --
Alternatively, the orientational vector $\mathbf{n}$ of a chiral active particle can be described through the chiral active Ornstein-Uhlenbeck model, as introduced in Ref.~\cite{caprini2019active}. In this case, the time evolution of $\mathbf{n}$ can be conveniently expressed in Cartesian coordinates, such that
\begin{equation}
\label{eq:AOUPdynamics}
\dot{\mathbf{n}}= - \frac{\mathbf{n}}{\tau} + \sqrt{\frac{1}{\tau}} \boldsymbol{w} + \Omega \,\hat{\mathbf{z}}\times \mathbf{n}  \,.
\end{equation}
Here, $\boldsymbol{w}$ is a white noise vector with zero average and unit variance, $\hat{\mathbf{z}}$ represents a unit vector normal to the plane of motion, and $\tau$ can be identified with the persistence time of the chiral active particle.
The last term in the right-hand side of Eq. \eqref{eq:AOUPdynamics}, $\Omega \hat{\mathbf{z}}\times \mathbf{n}$, acts similarly to a Lorentz force on the orientational vector $\mathbf{n}$ and, indeed, induces rotations in the particle orientation and trajectory. Again, this allows us to identify $\Omega$ as the particle chirality.

As previously discussed, the ABP and AOUP models give rise to consistent results if $D_r=1/\tau$. This mapping still holds in the presence of a non-vanishing chirality, and can be viewed by analyzing the autocorrelation of the orientational vector $\mathbf{n}$, which in both cases has an exponential shape~\cite{farage2015effective} (see Ref.~\cite{caprini2022parental} and Ref.~\cite{santra2025universal} for a direct comparison between the two models without and with chirality, respectively):
\begin{equation}
\langle \mathbf{n}(t)\cdot \mathbf{n}(0)\rangle = e^{ -\frac{t}{\tau}} \cos{\left(\Omega t \right)} \,,
\end{equation}
i.e.\ is characterized by an exponential decay with autocorrelation time $\tau$, modulated by periodic oscillation with frequency $\Omega$.
These oscillations account for the rotational motion of the single-particle dynamics and give rise to an oscillating mean-square displacement through the Kubo relation~\cite{van2008dynamics,caprini2019active}.

\subsection{Coarse-grained description for chirality-induced transverse forces}

\begin{figure}[!t]
\centering
\includegraphics[width=0.99\linewidth,keepaspectratio]{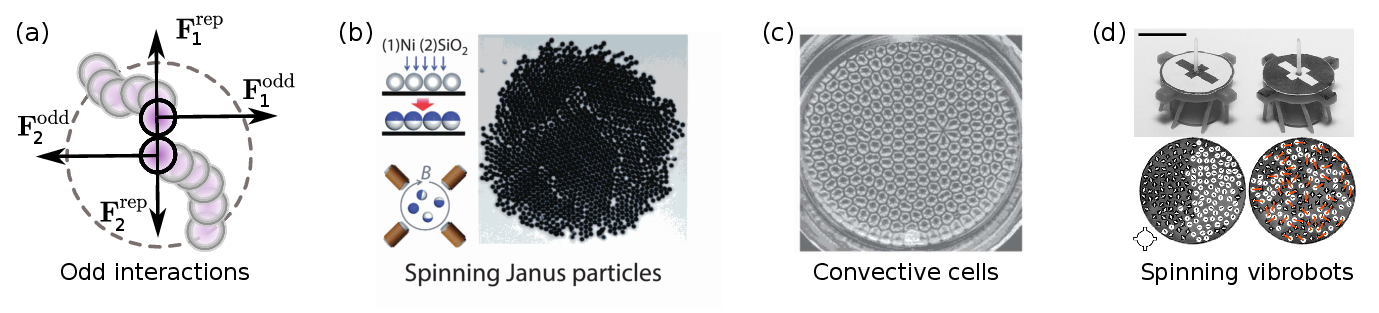}
\caption{\label{fig:fig1}
(a) Illustration of a collision between two chiral particles, governed by pure repulsive forces, $\mathbf{F}_i$ and odd interactions $\mathbf{F}^{odd}_i$.
The repulsive forces are directed tangentially along with the direction connecting the centers of the two particles, while odd interactions are directed normally compared to this direction.
Panels (a) is adapted with permission from Ref.~\cite{caprini2025bubble}; copyright (2025) AIP Publishing.
(b) Colloidal particles driven to spin by external magnetic fields. Panel (b) is adapted with permission from Ref.~\cite{yan2015rotating}.
(c) Rayleigh--Bénard convection cells organized in a hexagonal pattern, which show odd elasticity when the system is put under rotation. Panel adapted from Ref.~\cite{an2010int}.
(d) Left-handed and right-handed vibrobots manufactured by 3D printing. Panel (d) is adapted with permission from Ref.~\cite{scholz2021surfactants}.
}
\end{figure}

\noindent
{\it Conservative interactions} -- 
When two particles in equilibrium interact, they typically experience either a repulsive or an attractive force that can be derived from a potential.
For a particle $i$, the force due to the surrounding particles is given by $\mathbf{F}_i=- \nabla_i U_{tot}$, where $U_{tot}=\sum_{i<j} U(|\mathbf{x}_i -\mathbf{x}_j|)$,
  where , $U_{tot}$ denotes the total potential expressed as the sum of pairwise contributions $U(|\mathbf{x}_i -\mathbf{x}_j|)$ 
 that depend solely on the interparticle distance.
 This force is conservative and points along the line connecting the centres of the two particles (Fig. 1). Such interactions are often used to model volume exclusion effects, for example by employing the Weeks-Chandler-Andersen potential \cite{weeks1971role}.
  When two chiral active particles interact, conservative forces are included in their dynamics in the same way as for passive particles. For instance, when two chiral spinners come into contact, they are subject to volume exclusion effects; similarly, when two chiral active colloids approach each other, they may experience an effective attraction -- typically modelled by a Lennard-Jones potential -- which accounts for both van der Waals forces and hard-core repulsion.

\vskip10pt
\noindent
{\it Effective odd (transverse) interactions due to the particle chirality} --
Chiral active units experience additional effective forces, $\mathbf{F}^{odd}_i$, which act transversely to the line connecting the centres of two particles (Fig.~\ref{fig:fig1}(a)).
As a result, these interactions are non-conservative and cannot be derived from a potential. They can be written in two dimensions as
\begin{equation}
\label{eq:oddinteraction}
\mathbf{F}^{odd} = - \omega\nabla_i \sum_{i<j} U^{odd}(|\mathbf{x}_i - \mathbf{x}_j|) \times \hat{\mathbf{z}} \,,
\end{equation}
where $\hat{\mathbf{z}}$ is the unit vector normal to the $xy$ plane of motion, using a right-handed Cartesian coordinate system.
Here, $U^{odd}(|\mathbf{x}_i - \mathbf{x}_j|)$
 is a dimensionless function of the interparticle distance, whose form depends on the specific system and on the microscopic mechanism generating odd interactions. The strength of these interactions is determined by the parameter 
$\omega$, which encodes the particle chirality and will, in the following, be referred to simply as {\it chirality}.

Odd interactions can be understood as emerging from a coarse-graining procedure that takes place whenever two rotating objects come into direct or indirect contact -- for example, in granular spinners or in chiral colloids suspended in a fluid, where hydrodynamic interactions are significant.

\vskip10pt
\noindent
{\it Odd interactions for rotating objects in a fluid} -- A rotating object in a fluid -- for example, in a Newtonian fluid at low Reynolds number (Stokes regime) -- generates a rotating flow field around it. In the case of a sphere, this flow is known as a  {\it rotlet}  and decays as $1/r^2$, producing a circular velocity field around the particle.
When two rotating particles come close to each other, each experiences the flow field induced by the other. This induced flow is transverse to the line connecting the centres of the two particles, which means that the particles effectively interact through transverse (odd) forces of hydrodynamic origin.
Such hydrodynamic odd interactions arise in several systems: for example, in active colloids spinning under the influence of an external magnetic field~\cite{yan2015rotating,massana2021arrested} (Fig.\ref{fig:fig1}~(b)), in bacteria self-organizing into rotating crystals~\cite{petroff2015fast}, convective cells~\cite{an2010int} (Fig.\ref{fig:fig1}~(c)), or in more complex biological systems such as starfish embryos~\cite{tan2022odd}.

\vskip10pt
\noindent
{\it Odd interactions as a result of direct contact of rotating objects} --
When two granular spinners collide~\cite{scholz2018rotating,scholz2021surfactants,caprini2024spontaneous} (Fig.~\ref{fig:fig1}~(d)), the contact forces include both normal and tangential components. The normal force governs the repulsion, while the tangential force -- arising from rotational friction at the point of contact -- couples the angular motion to linear momentum exchange.
Because each spinner is rotating, the tangential frictional force during contact is systematically biased in a transverse direction, meaning that the impulse they impart on one another is not aligned with the line connecting their centres. This mechanism gives rise to effective transverse forces between the spinners: co-rotating particles can deflect sideways or even orbit around each other, while counter-rotating particles scatter more symmetrically.
Thus, rotational friction at the particle–particle contact provides the key microscopic mechanism underlying emergent transverse interactions in granular spinner systems.

\subsection{Odd interactions emerging in models for granular spinners}

More specifically, the odd interaction $\mathbf{F}^{odd}_i$ discussed in this paper can be regarded as a limiting case of a well-known class of dynamics in the study of granular spinners, namely inertial granular objects that undergo self-rotation due to a constant active torque.
This model was recently investigated by Digregorio, Pagonabarraga and Vega Reyes~\cite{digregorio2025phase}, who considered a system of $N$ two-dimensional rough disks of mass $m$, diameter $\sigma$, and moment of inertia $I$. Each disk undergoes translational motion with velocity $\mathbf{v}_i$ in the $x$,$y$ plane together with rotational dynamics with angular velocity $\boldsymbol{\omega}_i$ along the $z$ axis.
In particular, the center-of-mass velocities evolve according to a translational dynamics equivalent to Eq.~\eqref{eq:translationaldynamics}, subject to a frictional force, thermal noise, and inter-particle interactions, which embody the system activity.
Volume exclusion is enforced through normal forces, $\bm{F}_i^{\rm{n}}$, which can be derived from the gradient of a potential.
However, instead of odd interactions, the authors include tangential forces, $\bm{F}_i^{\rm{t}}$, as commonly done in models of granular particles. These tangential forces are given by
\begin{equation}
\bm{F}_i^{\rm{t}} = -\eta \sum_{j\neq i} \Theta(\sigma -x_{ij} )\,\Bigl( \bm{v}_{ij} - (\bm{v}_{ij} \cdot \hat{\bm{n}}_{ij})\hat{\bm{n}}_{ij} -\sigma \bm{\omega}_{ij} \times \hat{\bm{n}}_{ij}\Bigr) \,,
\label{tangential} 
\end{equation}
where $\bm{v}_{ij}=\bm{v}_j-\bm{v}_i$ denotes the relative velocity of the particle centers of mass and $x_{ij}=|\mathbf{x}_i -\mathbf{x}_j|$ their mutual separation.
The unit vector $\hat{\bm{n}}_{ij}=(\bm{x}_i-\bm{x}_j)/x_{ij}$ specifies the normalised separation direction, while $\Theta(\sigma-x_{ij})$ is  the Heaviside step function, restricting the range of this interaction at the particle diameter $\sigma$. Consequently, $\bm{F}_i^{\rm{t}}$ can be regarded as a contact force. 
Here, $\bm{\omega}_{ij}=(\bm{\omega}_i+\bm{\omega}_j)/2$ denotes the mean angular velocity of the two interacting particles, accounting for the particle rotations during collisions. This coupling between translational and angular velocity will prove to be fundamental in what follows.

Since the massive disks are not smooth and can rotate around their axes, their angular velocity $\bm{\omega}_i$ evolves through a rotational underdamped dynamics, subject to a friction with the medium, $-\gamma_{\theta}\boldsymbol{\omega}_i$ and a stochastic torque $\boldsymbol{\xi}_{\theta,i}$ with zero mean and correlations given by
 $\langle \boldsymbol{\xi}_{\theta,i}(t) \boldsymbol{\xi}_{\theta,j}(t^{\prime}) \rangle = 2\gamma_{\theta} T_{\rm{th}} \bm{1} \delta_{ij} \delta(t-t^{\prime})$, where $T_{th}$ sets the strength of the noise, $\gamma_{\theta}$ the rotational diffusion coefficient, and $\bm{1}$ the identity matrix.
The dynamics for $\boldsymbol{\omega}_i$ reads
\begin{equation}
   \label{eq:motion_rot}
    I \frac{d \boldsymbol{\omega}_i}{dt} =-\gamma_{\theta} \boldsymbol{\omega}_i + \boldsymbol{\xi}_{\theta,i} + \boldsymbol{\tau}_i + \boldsymbol{\tau}_0 \,,
\end{equation}
where the vector $\boldsymbol{\tau}_i$ is the torque generated by the tangential force and $\boldsymbol{\tau}_0$ is a constant torque which induce spinning trajectory and, thus, is responsible for the particle chirality.
Notice that the parity symmetry,  is broken by the term $\bm{\omega}_{ij} \times \hat{\bm{n}}_{ij}$ in Eq.~\eqref{tangential} which depends on $ \boldsymbol{\tau}_0$ and generates the handedness of the force $\bm{F}_i^{\rm{t}}$.

For large rotational friction $\gamma_{\theta}$ when the torque $\boldsymbol{\tau}_i$ can be neglected, the angular velocity is mainly determined by the constant torque $\boldsymbol{\tau}_0$, as 
\begin{equation}
\label{eq:over_dyn}
\gamma_{\theta} \bm{\omega}_i \approx \boldsymbol{\tau}_0 +  \boldsymbol{\xi}_{\theta,i}\,,
\end{equation}
corresponding to a chiral active Brownian motion in the overdamped limit equivalent to Eq.~\eqref{eq:chiralABP}.
By plugging the expression for $\boldsymbol{\omega}_{i}$ in the equation for the tangential force~\eqref{tangential}, we obtain
\begin{eqnarray}
\bm{F}_i^{\rm{t}} 
&&\approx -\eta \sum_{j\neq i} \Theta(\sigma -x_{ij} )\,\Bigl( \bm{v}_{ij} - (\bm{v}_{ij} \cdot \hat{\bm{n}}_{ij})\hat{\bm{n}}_{ij}  + \frac{\sigma(\boldsymbol{\xi}_{\theta,i}+\boldsymbol{\xi}_{\theta,j})}{2\gamma_\theta} \times \hat{\bm{n}}_{ij}\Bigr) + \mathbf{F}^{odd}_i \,.
\label{eq:Ft_approx}
\end{eqnarray}
Here, we have identified the term proportional to $\boldsymbol{\tau}_0=(0, 0, \tau_0)$, resulting from Eq.~\eqref{eq:over_dyn}, with the specific odd interaction form related to this model, given by
\begin{equation}
\mathbf{F}^{odd}_i = \eta \sum_{j\neq i} \Theta(\sigma -x_{ij} ) \frac{\sigma\boldsymbol{\tau}_{0}}{\gamma_\theta} \times \hat{\bm{n}}_{ij} \,.
\label{eq:eff_Fodd}
\end{equation}
The force~\eqref{eq:eff_Fodd} takes the form of Eq.~\eqref{eq:oddinteraction} because the cross product $\boldsymbol{\tau}_0 \times \hat{\mathbf{n}}_{ij}$ selects the direction perpendicular to the radial one. This means that the effective force~\eqref{eq:eff_Fodd} acts as transverse (odd) forces. In the granular spinners model, the odd interaction has a specific form  characterized by a linearly truncated profile $U^{odd}(\mathbf{x}_{ij})= (|\mathbf{x}_i - \mathbf{x}_j|-\sigma ) \,\Theta(\sigma - |\mathbf{x}_i - \mathbf{x}_j| )$ with an interaction strength given by $\omega=\eta \sigma \tau_0/\gamma_\theta$.

Within this approximation, the last term in Eq.~\eqref{eq:Ft_approx} acts as a stochastic noise which simply enhances the effective diffusion coefficient of the translational dynamics. By contrast, the first two terms promote alignment between the particle velocities since the quantity $\bm{v}_{ij} - (\bm{v}_{ij} \cdot \hat{\bm{n}}_{ij})\hat{\bm{n}}_{ij}$ corresponds to the transverse component of the relative velocity.
Therefore, under the identification in Eq.~\eqref{eq:Ft_approx}, a particle model governed exclusively by transverse forces can be interpreted as a limiting case of the granular spinners model, where the angular dynamics is dominated by a constant torque, and the effective force aligning the tangential components of the particle velocities can be neglected.

\section{A chirality-induced collective effect: bubbles induced by odd interactions -- the BIO phase}\label{sec:BIO}

Chiral active particles governed by odd (transverse) interactions exhibit novel collective phenomena compared to active matter systems that interact solely through conservative forces. In what follows, we consider the small-speed limit $v_0 \to 0$, where the active force can be neglected relative to odd interactions (transverse forces) and thermal noise. In this regime, the dynamics of the velocity $\mathbf{v}_i=\dot{\mathbf{x}}_i$ of the $i$-th particle are given by
\begin{equation}
\label{eq:dynamics_interacting}
m\dot{\mathbf{v}}_i =-\gamma \mathbf{v}_i +\gamma\sqrt{2D_t}\boldsymbol{\xi}_i + \mathbf{F}_i + \mathbf{F}^{odd}_i
\end{equation}
where we adopt the same notation as in Eq.~\eqref{eq:translationaldynamics}.
Here, $\mathbf{F}_i=-\nabla_i U_{\rm tot}$ is the conservative force arising from the total potential $U_{\rm tot}=\sum_{i<j} U(|\mathbf{x}_i - \mathbf{x}_j|)$, expressed as the sum of pairwise interactions. Since this term models steric repulsion between particles, we employ the Weeks–Chandler–Andersen potential, $U(r)=4\epsilon ((\sigma/r)^{12} - (\sigma/r)^6)$, 
where $\epsilon$ sets the energy scale and $\sigma$ denotes the nominal particle diameter.
The term $\mathbf{F}^{\rm odd}_i$ represents the odd interactions given by Eq.~\eqref{eq:oddinteraction}, which can be written as
$\mathbf{F}^{odd} = - \omega\nabla_i \sum_{i<j} U_{odd}(|\mathbf{x}_i - \mathbf{x}_j|) \times \hat{\mathbf{z}}$.
In Ref.~\cite{caprini2025bubble}, we chose $U_{\rm odd}(r)=\omega\sigma/r$ as a long-range interaction, inspired by the experimental study of colloidal spinners reported in Ref.~\cite{massana2021arrested}.
We stress that even if the particle is not polar, the condition $v_0 \to 0$ does not correspond to a passive limit. Indeed, transverse forces are activity-induced and emerge from the coupling between translational and rotational motion.


\subsection{The BIO phase}

\begin{figure}[!t]
\centering
\includegraphics[width=0.99\linewidth,keepaspectratio]{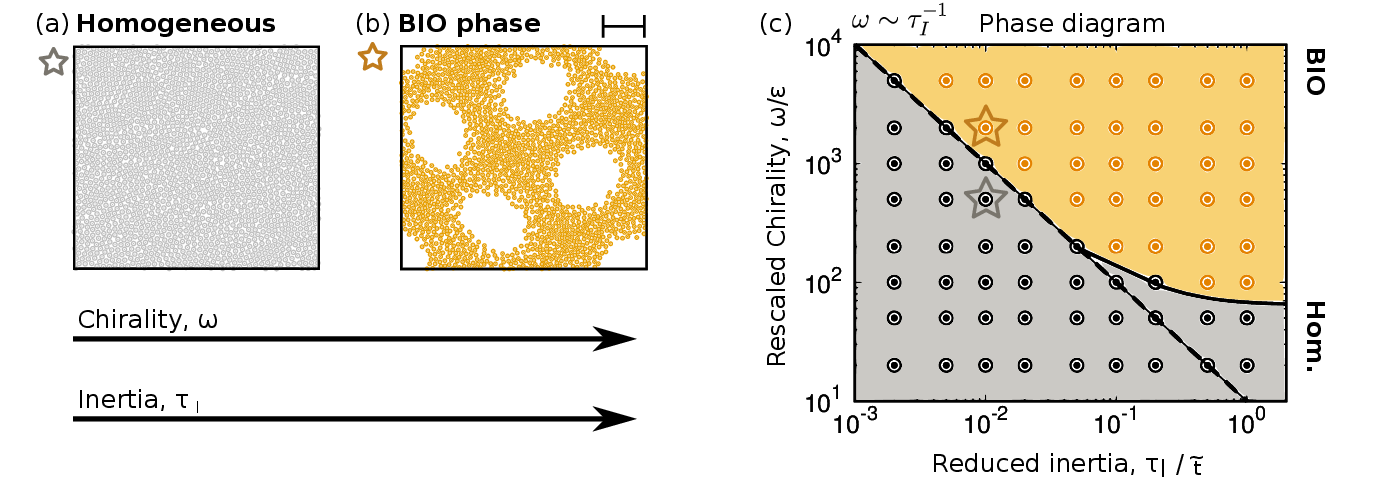}
\caption{\label{fig:fig2}
{\textbf{BIO phase: Bubbles induced by odd interactions.}} 
(a)-(b) Snapshot configurations showing the homogeneous and BIO (bubbles induced by odd interactions) phases. The scale bars referring to both panels denote 10 $\sigma$ where $\sigma$ is the particle diameter.
(c) Phase diagram in the plane of chirality $\omega/\epsilon$, i.e.\ the strength of odd interactions, and reduced inertia, $\tau_I/\tilde{t}$ (with $\tilde{t}=\sigma \sqrt{m/\epsilon}$), showing homogeneous (grey points) and BIO phases (yellow points). The dashed black line marks the scaling $\omega\sim1/\tau_I$ for the transition line from the two phases, while the two stars correspond to the two configurations in panels (a) and (b). Panels (a), (b), and (c) are adapted with permission from Ref.~\cite{caprini2025bubble}; copyright (2025) AIP Publishing.
}
\end{figure}

When odd interactions are strong (see Eq.~\eqref{eq:oddinteraction}), i.e., in the regime of large chirality $\omega$, the system spontaneously evolves from a homogeneous state (Fig.\ref{fig:fig2}(a)) to an inhomogeneous phase (Fig.~\ref{fig:fig2}~(b)), characterized by the emergence of bubbles, as reported in Ref.~\cite{caprini2025bubble}. These bubbles are devoid of particles and, in the steady state, arrange themselves preferentially into an almost regular lattice with hexagonal order.
This new collective phenomenon is termed BIO (bubbles induced by odd interactions) and constitutes a novel phase uniquely driven by odd (transverse) interactions, or equivalently, by chirality. Importantly, the BIO phase emerges spontaneously, i.e., in the absence of attractive interactions, and is fundamentally distinct from phase separation. In particular, the bubbles acquire a finite steady-state size, meaning that the coarsening process of bubble formation is interrupted.

By monitoring the steady-state kinetic energy and the average vorticity as order parameters, Ref.~\cite{caprini2025bubble} demonstrates that chirality drives a genuine non-equilibrium phase transition from the homogeneous state to the BIO phase. This analysis identifies a critical value of the chirality, $\omega_c$, which in general depends on the inertial time.
The collective behavior is systematically explored through a phase diagram in the plane defined by chirality $\omega$ and inertial time $\tau_I=m/\gamma$ (Fig.~\ref{fig:fig2}(c)). 
Both $\omega$ and $\tau_I$ favor the BIO phase, giving rise to a homogeneous–BIO transition line that scales as
\begin{equation}
\label{eq:scaling}
\omega_c \sim \frac{1}{\tau_I} \, ,
\end{equation}
as shown numerically in Fig.~\ref{fig:fig2}~(c). Thus, the larger the inertial time, the smaller the critical chirality required to trigger the BIO phase.
This numerical analysis reveals that inertia is essential for the emergence of the BIO phase, which is suppressed in the overdamped limit $\tau_I \to 0$, where the system remains homogeneous.

\begin{figure}[!t]
\centering
\includegraphics[width=0.99\linewidth,keepaspectratio]{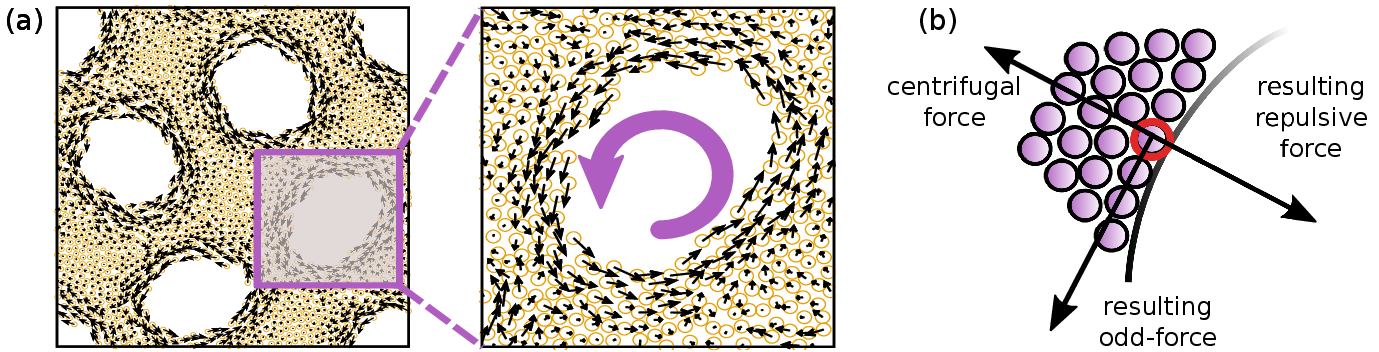}
\caption{\label{fig:fig3}
{\textbf{Edge currents and centrifugal forces generated by odd interactions.}}
(a) Snapshot configuration showing a BIO phase (bubbles induced by odd interactions) where black arrows denote the particle velocities and points represent particles. In addition, a zoom of panel (a) is reported with an additional violet arrow showing edge currents (counterclockwise rotation). (b) illustration of the mechanism responsible for the bubble stability showing the competition between the net force, due to repulsive interactions, and the centrifugal force generated by odd interactions.
Panels (a) and (b) are adapted with permission from Ref. \cite{caprini2025bubble}; copyright (2025) AIP Publishing.
}
\end{figure}

\vskip10pt
\noindent
{\it Physical origin of the BIO phase: a scaling argument} --
To explain the spontaneous emergence of bubbles, it is useful to first examine the system dynamics in the BIO phase, specifically by representing particle velocities as black arrows (Fig.~\ref{fig:fig3}(a)). While particles in the bulk display nearly random velocities, those located at the bubble interface move coherently, tangentially to the bubble surface.
These coherent rotations are reminiscent of the edge currents that characterize chiral active particles. This observation is consistent with the findings of Caporusso, Gonnella, and Levis~\cite{caporusso2024phase}, who reported edge currents at the boundaries of clusters formed by overdamped chiral particles subject to transverse forces and strong attractions.
Edge currents emerge whenever the system exhibits density inhomogeneity, as occurs at the bubble interface. This mechanism can be understood through a simple force-balance argument on a test particle (Fig.\ref{fig:fig3}(b)). A particle located in the bulk, isotropically surrounded by other chiral units, experiences a vanishing average force and thus carries no net momentum. By contrast, a particle at the bubble surface is subject to two distinct net forces.
The first is directed radially inward toward the bubble center and arises from repulsive interactions mediated by the WCA potential. This contribution provides the standard pressure that tends to suppress bubbles and smooth out density inhomogeneities, as expected in equilibrium. The second is directed tangentially along the bubble interface and originates from the unbalanced components of odd interactions exerted by surrounding particles. This force sustains the edge currents, generating a finite momentum flux tangential to the bubble surface.
Bubbles form and remain stable because a rotating particle at their interface also experiences a centrifugal force pointing radially outward. 
Thus, in the presence of edge currents, a bubble can stabilize at a given radius when its centrifugal force balances the inward force due to repulsive interactions.

\vskip10pt
\noindent
{\it Spatial velocity correlations as a signature of the homogeneous-BIO phase transition} --
The kinetic energy and average vorticity fields exhibit a sharp increase as a function of chirality at the homogeneous–BIO phase transition. These observables can therefore be regarded as order parameters, suggesting that the phenomenon can indeed be interpreted as a non-equilibrium phase transition.
To further support this interpretation, we analyze the steady-state spatial velocity correlations~\cite{caprini2025odd} in the homogeneous regime, i.e., prior to the onset of the BIO phase. For analytical tractability, this study is performed in a high-density system exhibiting solid-like order, for example with hexagonal symmetry (triangular lattice). In this case, both the repulsive and odd interactions can be linearized, so that the dynamics \eqref{eq:dynamics_interacting} reduce to the following equation of motion:
\begin{equation}
m \dot{\mathbf{v}}_i= - \gamma \mathbf{v}_i + C_0\sum_\mb^{n.n}\ (\bu_\mb-\bu_\bn)
+C_1\sum_\mb^{n.n} {\hat z} \times(\bu_\mb-\bu_\bn) \,. 
\end{equation}
Here, $\sum_\mb^{n.n}$ denotes the sum over nearest neighbors (first shell). The coefficient $C_0$ is approximately given by the second derivative of the potential evaluated at the lattice spacing, corresponding to the second-order term in the Taylor expansion of the interaction potential $U$.
The coefficient $C_1$ is related to the first derivative of the odd interactions and, like $C_0$, depends on the details of the lattice. While it is reasonable to restrict the Taylor expansion of the short-ranged WCA potential to the first shell of neighbors, this represents a rough approximation for the odd force, where contributions from further shells should also be taken into account. Here, however, we restrict the analysis to the first shell for simplicity, while a more complete numerical study including higher-order neighbors is reported in Ref.~\cite{caprini2025odd}.
The linearization of the force—valid for an ideal crystal—allows us to predict analytically the spatial velocity correlations in Fourier space. These correlations exhibit a maximum at a finite value of the reciprocal vector $\mathbf{q}$, denoted $\mathbf{q}^*$, and can be approximated as
\begin{equation}
\label{eq:spatial_velocity_correlation}
\langle \hat{\mathbf{v}}(\mathbf{q}) \hat{\mathbf{v}}(-\mathbf{q})\rangle 
\approx \frac{2T}{m} \left(1 + C(\mathbf{q}) \right)
\end{equation}
where $\hat{\mathbf{v}}(\mathbf{q})$ is the Fourier transform of the velocity field and the function $C(\mathbf{q})$ reads
\begin{equation}
C(\mathbf{q}) \propto \frac{1}{a+b (\mathbf{q}-\mathbf{q}^* )^2} \,.
\end{equation}
The constants $a$ and $b$ depend on the model parameters, specifically on the inertial time $\tau_I=m/\gamma$ and the effective spring constants $C_0$ and $C_1$:
\begin{eqnarray}
&b=\beta_2 \tau_I^2 \frac{C_1^2}{m C_0}\\
&a=1-\beta_1\tau_I^2 \frac{C_1^2}{m C_0} \,.
\end{eqnarray}
Here, $\beta_1$ and $\beta_2$ are numerical coefficients determined by the lattice geometry.
The explicit expressions for these terms depend on the details of the interactions. In general, however, $C_0$ scales with the energy scale of the WCA potential ($C_0 \sim \epsilon/\sigma^2$), while $C_1$ is proportional to the strength of odd interactions ($C_1 \sim \omega/\sigma^2$). Finally, the characteristic wavevector $\mathbf{q}^*$ depends on lattice properties and, in particular, on the number of shells included in the linear approximation of transverse forces $\mathbf{F}^{odd}$ (see Ref.~\cite{caprini2025odd} for details).
The prediction~\eqref{eq:spatial_velocity_correlation} implies that the system spontaneously develops spatial correlations in the velocity field, as evident from the $\mathbf{q}$-dependence of the second term. These correlations are a hallmark of the system’s non-equilibrium nature and exhibit a divergence at a finite wavevector $\mathbf{q}=\mathbf{q}^*$, for parameter values such that $a=0$. In other words, the instability condition reads
\begin{equation}
\label{eq:relation_instability}
m C_0 = \beta_1 \tau_I^2 C_1^2 \,.
\end{equation}
This relation determines the stability threshold of a chiral active system governed by transverse interactions and reproduces the homogeneous–BIO transition line numerically observed in Ref.~\cite{caprini2025bubble} within the phase diagram spanned by the inertial time $\tau_I$ and chirality $\omega$. In particular, the critical chirality $\omega_c$ depends on the inertial time as
\begin{equation}
\omega_c^2 \sim \frac{m \epsilon}{\tau_I^2 \sigma^2} \,,
\end{equation}
which recovers the scaling relation \eqref{eq:scaling} obtained numerically. Thus, the larger the inertial time, the smaller the critical chirality $\omega_c$ required for the emergence of the BIO phase.
Finally, taking the inverse Fourier transform of Eq.~\eqref{eq:spatial_velocity_correlation}, we obtain an oscillatory profile for the spatial velocity correlations in real space:
\begin{equation}
\label{eq:spatialvelocitycorrelations}
\langle \mathbf{v}(r)\mathbf{v}(0)\rangle \propto \cos{(q^* r)} \frac{e^{-r /\xi}}{r^{1/2}}
\end{equation}
where $q^*=|\mathbf{q}^*|$ and $\xi$ corresponds to the correlation length of the spatial velocity correlations with the following expression
\begin{equation}
\xi = \sqrt{\frac{\beta_2 \tau_I^2 \frac{C_1^2}{m C_0}}{1-\beta_1\tau_I^2 \frac{C_1^2}{m C_0}}} \,.
\end{equation}
The profile in Eq.~\eqref{eq:spatialvelocitycorrelations} implies that spatial velocity correlations decay exponentially while exhibiting oscillatory behavior. Consequently, the system is characterized by vortex structures with a typical size of order $\sim 1/q^*$.
We emphasize that the prediction for the correlation length $\xi$ suggests the occurrence of a second-order phase transition, due to the divergent behavior of $\xi$ when the instability condition in Eq.~\eqref{eq:relation_instability} is satisfied.

\section{Conclusions and perspectives}\label{sec:conclusion}

In this paper, we have reviewed the theoretical models and emergent phenomena characteristic of chiral active systems.
The single-particle dynamics of circular microswimmers or rotating objects can be described by chiral active Brownian or chiral active Ornstein–Uhlenbeck models, which reproduce circular self-propelled motion. We have discussed in detail the effective interactions that govern the dynamics of chiral systems.
Chiral particles not only interact through standard conservative forces, accounting for volume exclusion effects or attractions, as in equilibrium systems,
but are also subject to additional effective forces -- termed odd interactions -- arising from a coarse-graining of the physical mechanisms that govern their dynamics, such as the hydrodynamic flow generated by rotating objects or the rotational friction in macroscopic granular spinners. These odd interactions break time-reversal symmetry, are intrinsically non-conservative, and transfer angular momentum, thereby inducing rotations in particle motion.

For more than a decade, it has been known that active matter exhibits collective phenomena, generally classified into two main types of non-equilibrium phase transitions.
Non-chiral active particles, characterized by persistent self-propelled motion, undergo motility-induced phase separation (MIPS)~\cite{cates2015motility}, i.e., they exhibit phase coexistence between dense clusters and a dilute gas even in the absence of attractive forces~\cite{fily2012athermal,buttinoni2013dynamical,solon2015pressure,digregorio2018full,mandal2019motility,caprini2020spontaneous,omar2023mechanical}.
Active systems with alignment mechanisms -- whether orientation-orientation or velocity-orientation -- display flocking behavior~\cite{vicsek2012collective,cavagna2014bird,barberis2016large,shankar2017topological,caprini2023flocking}, forming polarized homogeneous phases or band-like structures that move coherently in a single direction~\cite{ihle2011kinetic,solon2015phase,giavazzi2018flocking,das2024flocking,musacchio2025self,casiulis2024geometric}.
Our recent studies expand this scenario by introducing a novel collective phenomenon, uniquely induced by odd (transverse) interactions due to chirality. Chiral systems undergo a transition from a homogeneous phase to an inhomogeneous phase characterized by bubble-like structures. We term this state the BIO phase -- bubbles induced by odd interactions~\cite{caprini2025bubble}. These bubbles exhibit edge currents at their interfaces and reach a steady-state size when the centrifugal force generated by odd interactions balances the repulsive forces. This mechanism, along with the resulting interruption of coarsening, distinguishes the BIO phase from the well-known MIPS scenario.

Future directions for the study of odd-interacting chiral matter include the theoretical understanding and characterization of the BIO phase within the framework of non-equilibrium critical phenomena. Key open questions involve the role of density in stabilizing the BIO phase, the scaling of bubble size, and the order of the homogeneous-to-BIO transition, which requires further numerical investigation beyond the preliminary arguments presented here.
From a theoretical standpoint, the BIO phase still lacks a systematic description in terms of hydrodynamic or kinetic theory, for instance via a Boltzmann-equation framework. Another open challenge is the development of a scalar field theory capable of reproducing the bubble phase reported here. Although a scalar field theory for chiral systems has recently been proposed in Ref.~\cite{huang2025anomalous}, it cannot capture the BIO phase described in this work, as it is derived from overdamped dynamics.
In summary, the BIO phase defines a novel class of collective phenomena in the active matter paradigm, raising  several open questions
and opportunities for both theoretical  and experimental investigation.

\section{Acknowledgement}

We thank Pasquale Digregorio, Ignacio Pagonabarraga, and Francisco Vega Reyes for useful discussions.
This paper is based on the talk presented by LC at the Conference StatPhys29 -- Florence 2025 -- in occasion of the early career award in Statistical Physics.
LC acknowledges financial support from the University of Rome Sapienza, under the project Ateneo 2024 (RM124190C54BE48D) ``Elementary excitations at the origin of glassy or hexatic behavior in low dimensional system, at and out of equilibrium''.

--



\section*{References}

\bibliography{Respbib.bib}

\end{document}